\documentclass[aps,prd,groupedaddress,showpacs,slide]
          {revtex4}%
\usepackage{epsfig}
\usepackage{rotating}
\usepackage{array}
\usepackage{epsfig}
\usepackage{graphicx}
\usepackage{epstopdf}
\usepackage{amssymb}
\usepackage{latexsym}

\newcommand{\be}{\begin{eqnarray}}
\newcommand{\bb}{\bibitem}
\newcommand{\ee}{\end{eqnarray}}

\newcommand{\fig}{\begin{figure}}
\newcommand{\ef}{\end{figure}}
\newcommand{\bc}{\begin{center}}
\newcommand{\ec}{\end{center}}

\newcommand{\hs}{\hspace*{0.2in}}
\newcommand{\bn}{\begin{enumerate}}
\newcommand{\en}{\end{enumerate}}

\newcommand{\bz}{\begin{itemize}}
\newcommand{\ez}{\end{itemize}}
\newcommand{\ct}{\centerline}
\newcommand{\ep}{\epsfig}
\newcommand{\cp}{\caption}

\newcommand{\ba}{\begin{array}} 
\newcommand{\ea}{\end{array}}

\newcommand{\bt}{\begin{tabular}}
\newcommand{\et}{\end{tabular}}

\newcommand{\mc}{\mathcal}

\newcommand{\bd}{\begin{displaymath}}
\newcommand{\ed}{\end{displaymath}}
\newcommand{\nn}{\nonumber}
\newcommand{\ben}{\begin{eqnarray*}}
\newcommand{\een}{\end{eqnarray*}}
\newcommand{\mkb}{\makebox}

\newcommand{\al}{\alpha}

\newcommand{\bq}{\begin{quote}}
\newcommand{\eq}{\end{quote}}

\newcommand{\gsim}{\gtrsim}
\newcommand{\lsim}{\lesssim}
\begin{document}
\def\bea{\begin{eqnarray}}
\def\eea{\end{eqnarray}}
\def\nn{\nonumber}
\newcommand{\snu}{\tilde \nu}
\newcommand{\sll}{\tilde{l}}
\newcommand{\asnu}{\bar{\tilde \nu}}
\newcommand{\stau}{\tilde \tau}
\newcommand{\dmsnu}{{\mbox{$\Delta m_{\tilde \nu}$}}}
\newcommand{\mt}{{\mbox{$\tilde m$}}}

\renewcommand\epsilon{\varepsilon}
\def\be{\begin{eqnarray}}
\def\ee{\end{eqnarray}}
\def\lla{\left\langle}
\def\rra{\right\rangle}
\def\za{\alpha}
\def\zb{\beta}
\def\lsim{\mathrel{\raise.3ex\hbox{$<$\kern-.75em\lower1ex\hbox{$\sim$}}} }
\def\gsim{\mathrel{\raise.3ex\hbox{$>$\kern-.75em\lower1ex\hbox{$\sim$}}} }
\newcommand{\Rbs}{\mbox{${{\scriptstyle \not}{\scriptscriptstyle R}}$}}

\draft
\preprint{IPMU09-0106}


\title{Dark Matter Signals and Cosmic Ray Anomalies in an Extended Seesaw Model}


\thispagestyle{empty}
\author{ H. Sung Cheon$^{1,}$\footnote{E-mail:
        hscheon@gmail.com},~~ Sin Kyu Kang$^{2,}$\footnote{E-mail:
        skkang@snut.ac.kr},~~ C. S. Kim$^{1,}$\footnote{E-mail:
        cskim@yonsei.ac.kr,~~~ Corresponding Author} }
\affiliation{$ ^{1}$ Department of Physics and IPAP, Yonsei University, Seoul 120-749, Korea\\
             $ ^{2}$ School of Liberal Arts, Seoul National University of Technology, Seoul 121-742, Korea}

\begin{abstract}
\noindent

\noindent
An extended seesaw model  proposed to achieve low scale leptogenesis can
resolve the excess positron and electron fluxes observed from PAMELA, ATIC and/or Fermi-LAT, and
{\it simultaneously} accommodate some of recent experimental results for dark matter (DM) signals.
In this approach, in addition to $SU(2)_L$ doublet and the (light) singlet Higgs fields,
an extra vector-like singlet neutrino and a singlet scalar field,
which are coexisting two-particle dark matter candidates,
are responsible for the origin of the excess positron and electron fluxes
to resolve the PAMELA, ATIC and/or Fermi-LAT anomalies,
as well as for the DM signals observed from direct searches in low mass scale.

\end{abstract}

\maketitle \thispagestyle{empty}


\noindent{\bf I. Introduction:}
\\

The quest for identification of the missing mass of our universe is one of the most fundamental issue in astroparticle physics and cosmology.
The evidence for non-baryonic dark matter (DM) inferred from a combination of cosmological and astrophysical phenomena becomes more and more convincing,
which alludes the existence of new physics beyond the standard model (SM).
Very recently, several new exciting data on DM have been released, which may open up new era to search for DM in a low mass region
of a few GeV.  CDMS II collaboration reported the two DM candidate events with a 77\% C.L. and the upper bound of null result \cite{CDMS}.
DAMA collaboration confirmed the model independent evidence of the presence of DM on the basis of the DM annual modulation signature with 8.9$\sigma$
significance \cite{DAMA10}. The CoGeNT experiment reported a possible signal of a light DM candidate with $m_{\rm DM}=7$-$11$ GeV,
and provided 90\% C.L. WIMP exclusion plots as well \cite{cogent}.
Those three independent experimental results may be interpreted as signals of the existence of DM with a low mass around a few GeV \cite{scopel2}.
Contrary to  the results from CDMS II, DAMA and CoGeNT,  XENON100 collaboration announced that they have not observed any DM signal
for the similar parameter ranges searched by those three experiments \cite{XENON}.
Therefore, we need further experimental results to judge if there really exists a DM candidate with a low mass or not.

On the other hand, the PAMELA experiment has presented a significant positron flux excess over the expected background
with no excess in the corresponding anti-proton flux \cite{PAMELA}.
The ATIC/PPB-BETS experiment has shown significant excess of electron and positron flux at energies around 300-800 GeV \cite{ATIC,PPB}.
More recently, Fermi-LAT experiment have also shown an excessive electron and positron flux in the same energy range as in ATIC
 but its strength was not strong compared to ATIC \cite{Fermi}.
So, it is likely that  the experimental evidences for the signals of DM with a low mass scale
are not reconciled with the cosmic ray positron and electron excess
in the framework of one and only one DM scenarios.

Recently, we have proposed an extended seesaw model to simultaneously and naturally accommodate tiny neutrino masses, low scale leptogenesis
and dark matter candidate by introducing extra singlet neutrinos and singlet scalar particles on top of the canonical seesaw model  \cite{kk,ckk1}.
Furthermore,  we have proposed a coexisting two-particle DM scenario \cite{DDM}
by allowing both an extra singlet Majorana neutrino
and a light singlet scalar particle  as two DM candidates.
Such a scenario containing more than one DM may be desirable in the case that there exist a few
incompatible phenomena which are very hard to reconcile in the scenarios with only one DM.\\
\hs The purpose of this letter is to investigate how both the low mass DM signals observed from
direct DM searches and the cosmic ray positron and electron excess observed from PAMELA, ATIC and/or Fermi-LAT experiments are
simultaneously explained in the extended seesaw model with coexisting two-particle DM proposed in \cite{DDM}.
Due to the tension among the experimental results of direct search for DM in low mass scale, we first consider the case
that  lighter DM candidate in our model has mass around 3 GeV allowed by DAMA experiment, which is not in conflict with other null results
from direct searches but is inconsistent with the DM signals observed from CoGeNT.
The other case we consider is to accept DAMA and CoGeNT signals for DM candidate whose overlapped mass range lies
between  $7~\mbox{GeV}$ and $11~\mbox{GeV}$ while ignoring XENON100 results.
In this work, we slightly modify the model  proposed in \cite{DDM}
by replacing extra singlet Majorana neutrino with singlet vector-like neutrinos
so as to simply resolve the cosmic ray anomaly  while keeping to  accommodate  tiny neutrino masses
and low scale leptogenesis of order 1-10 TeV\cite{kk,ckk1}.

We notice that to achieve our coexisting two-particle DM scenario in the renormalizable
framework as shown in \cite{DDM},
an extra singlet Higgs scalar field $\Phi$ is necessarily introduced,
which may  open up new channels of DM annihilations.
As will be shown later, in this scenario, this scalar field $\Phi$ may play an
 essential role in resolving the unexpected electron and positron fluxes
 measured at PAMELA, ATIC and/or Fermi-LAT if the mass of $\Phi$ has rather small around just below
1 GeV so as for the annihilation cross section to be enhanced via
a mechanism first described by Sommerfeld \cite{Sommerfeld,XDM,Arkani}.
Once this new force carrier $\Phi$ is included, the possibility of a new dominant annihilation
of singlet vector-like neutrinos into a fair of $\Phi$ opens up.
The $\Phi$ mixes with the Higgs allowing it to decay into the final state fermions,
and if the $\Phi$ is taken to be light,
it is kinematically constrained to decay to mostly lepton pairs preventing from producing anti-protons,
so that the excess of positron and/or electron observed can be accounted for.
In addition, the low mass DM signals will be explained by considering the singlet scalar $\psi$
as the lightest DM candidate with mass
of order a few GeV. Thus, the low mass DM signals, the excess  positron and  electron fluxes produced from the cosmic rays,
low scale leptogenesis and light neutrino masses can be {\it simultaneously} accommodated in our model proposed.

To see how the coexisting two-particle DM scenario is achieved, let us consider the following  Lagrangian
\be
\mc{L}&=& \mc{L}_{\rm 0} +(Y_D \bar{L} H N + Y_S \bar{N} \psi S +h.c.)+ M_N N^{T}N
          + Y_\Phi \bar{S} \Phi S- m_{_S^0} \bar{S}S  \nn \\
          &+&\frac{1}{2}m^2 _{\psi^0} \psi^2
 -  \frac{\lambda_s}{4} \psi^4 - \lambda H^\dagger H \psi^2
 +  \frac{1}{2}m^2 _{\Phi^0} \Phi^2-\frac{\lambda_2}{4} \Phi^4
-\lambda_3 \psi^2 \Phi^2 - \lambda_4 H^\dagger H \Phi^2 ,\label{lag1}
\ee
where the first term is the Lagrangian of the SM and kinetic terms of the singlet fields,
and $L$, $N$, $S$ and $\psi$ stand for $SU(2)_L$ lepton doublet,
singlet heavy Majorana neutrino, singlet vector-like neutrino composed of two Weyl fermions,
and light singlet scalar, respectively.
Note that $S$ and $\psi$ are our coexisting two-particle dark matter candidates.
Finally $H$ and $\Phi$ denote the $SU(2)_L$ doublet and singlet (Higgs) scalar fields,
and $m_\Phi$  is assumed to be smaller than $1$ GeV to realize the Sommerfeld enhancement
in indirect detection \cite{Sommerfeld, XDM, Arkani}.
The effective scalar potential including one-loop corrections is given by
\be V_{eff} &=& -\frac{1}{2}m^2 _\psi \psi^2 +\frac{\lambda_s}{4}\psi^4
 + \lambda H^\dagger H \psi^2 -\frac{1}{2}m^2 _H  H^\dagger H
+ \frac{\lambda_1}{4}H^\dagger H H^\dagger H - \frac{1}{2} m^2 _\Phi \Phi^2
 + \frac{\lambda_2}{4}\Phi^4 + \lambda_3 \psi^2 \Phi^2 \nn\\
&+&  \lambda_4 H^\dagger H\Phi^2 + \frac{1}{64\pi^2}\Big [m^4 _H
 \Big ( \ln\frac{m^2 _H}{\mu^2} - \frac{3}{2} \Big )
+ 2m^4 _Z \Big ( \ln \frac{m^2 _Z}{\mu^2} - \frac{5}{6} \Big )
+ 4m^4 _W \Big (\ln\frac{m^2 _W}{\mu^2} - \frac{5}{6} \Big ) \nn \\
&-& 12 m^4 _t \Big ( \ln \frac{m^2 _t}{\mu^2} - \frac{3}{2} \Big )
+ m^4 _\psi \Big ( \ln\frac{m^2 _\psi}{\mu^2} - \frac{3}{2} \Big )
+  m^4 _\Phi \Big ( \ln \frac{m^2_\Phi}{\mu^2} - \frac{3}{2} \Big )
-4m^4 _S \Big (\ln\frac{m^2 _S}{\mu^2} - \frac{3}{2}\Big )\Big ], \label{eff}
\ee
where we have adopted $\overline{MS}$ renormalization scheme and  the field-dependent masses are
\begin{eqnarray*}
&&m^2 _t = y^2 _t h^2 /2, ~~~ m_Z ^2 = (g^2 + g'^2)h^2/4, ~~~ m^2 _W = g^2 h^2 /4, \\
&&m^2 _\psi = m^2 _{\psi^0} - 2 \lambda H^\dagger H -2 \lambda_3 \Phi^2, \\
&&m^2 _\Phi =m^2 _{\Phi^0}-\lambda_2 \Phi^2 -2 \lambda_4 H^\dagger H, \\
&&m^2 _H =m^2 _{H^0} - \lambda_1 H^\dagger H -2 \lambda_4 \Phi^2, ~~ m^2 _S = Y^2 _\Phi \Phi^2,
\end{eqnarray*}
where $\sqrt{2}H^T = (h,0)$.
In order to guarantee the stability of the 2DM candidates, we impose the discrete symmetry $Z_2 \times Z_2^{\prime}$
under which all SM bosons (photon, Higgs, $\mbox{W}^\pm$ and $Z$) and $\Phi$ are $(+,+)$, all SM fermions are $(-,-)$, the singlet neutrino $S$ is $(-,+)$ and the singlet scalar boson $\psi$ is $(+,-)$.
Now, we demand that the minimum of the scalar potential is bounded from below so
as to guarantee the existence of vacuum
and the minimum of the scalar potential must spontaneously
break the electroweak gauge group, $<H^0>, <\Phi> \neq 0$, but must not break
$Z_2 \times Z_2 '$ symmetry imposed above.

Since Eq. (\ref{eff}) depends on the renormalization scale $\mu$, it must be RG-improved and this can be simply done by repeatedly
decoupling all singlet particles and top quark at their mass scales \cite{raidal}.
 After spontaneous symmetry breaking, the low energy effective scalar potential becomes
\be V_{eff} &=&-\frac{1}{2}\bar{ m}^2 _\psi \psi^2 - \frac{1}{2} \bar{m}^2 _{h} h^2
 -\frac{1}{2}\bar{m}^2 _\phi \phi^2 + 2 \bar{\lambda}_4 v_h v_\phi h\phi
+ \frac{\lambda_s}{4} \psi^4 +\frac{\bar{\lambda}_1}{4}v_h h^3+\frac{\bar{\lambda}_1}{16} h^4
+ \frac{\bar{\lambda}_2}{4} \phi^4 \nn \\
&+& \bar{\lambda}_2 v_\phi \phi^3 + \frac{\lambda}{2}\psi^2 h^2 + \lambda v_h h \psi^2
 + \lambda_3 \psi^2 \phi^2 +2 \lambda_3 v_\phi \phi \psi^2
+\frac{\bar{\lambda}_4}{2}h^2\phi^2+\bar{\lambda}_4 v_\phi h^2 \phi
 + \bar{\lambda}_4 v_h h\phi^2 + h.c., \label{pot}
\ee
where $\bar{m}^2 _\psi = m^2_{\psi^0} +\lambda v^2 _h + 2 \lambda_3 v^2 _{\phi},~
\bar{m}^2 _h = \frac{1}{2}m^2 _{H^0} -\frac{3}{4} \bar{\lambda}_1v^2 _h -\bar{\lambda}_4 v_\phi ^2,~
\bar{m}^2 _\phi = m^2 _{\phi^0} - 3 \bar{\lambda}_2 v^2 _\phi - \bar{\lambda}_4 v^2 _h $.
Here, we have shifted the Higgs boson $H$
and the singlet Higgs scalar  $\Phi$ by $H\rightarrow h+v_h $ and $\Phi \rightarrow \phi + v_\phi$, respectively, and
\ben
\bar{\lambda_1} &=& \lambda_1 - \frac{3}{32\pi^2} \lambda^2 _1 + \frac{9}{32 \pi^2} y^4 _t - \frac{3}{8\pi^2}\lambda^2 -\frac{3}{8\pi^2}\lambda^2 _4,  \\
\bar{\lambda}_2 &=& \lambda_2 - \frac{3}{32\pi^2} (4 \lambda^2 _4 + 4 \lambda^2 _3 + \lambda^2 _2 - 4 Y^4 _\Phi), \\
{\rm and}~~~~\bar{\lambda}_4 &=& \lambda_4 - \frac{3}{128\pi^2}(4 \lambda_4 \lambda_1 + 8 \lambda \lambda_3 + 4 \lambda_2 \lambda_4).
\een
Since there exists a mixing mass term between $h$ and $\phi$,
we rotate them with $\phi =sh'+ c\phi'$ and $h=ch'-s\phi'$,
where $s$ and $c$ are  $\sin\theta$ and $\cos\theta$, respectively.

For $m_\phi \lsim 1$ GeV and $m_{_S} >> m_\phi$,  the singlet neutrinos $S$  annihilate into mostly $\phi \phi$.
Other annihilation channel like $\bar{S}S \rightarrow \psi\psi$ is negligible due to its very small coupling of the process.
The $\phi$'s can then subsequently decay into SM particles, which arises due to their mixing with the Higgs field $h$.
For the case of $m_\phi=0.25$ GeV, the $\phi$ mostly decays to muon pairs, which in turn produce electrons and positrons,
and thus the resulting spectra for the electrons and positrons are much harder than typical $e^{+}e^{-}$ spectra coming from
weak-scale WIMP annihilation as shown in \cite{XDM, Arkani}.

The amount of cold dark matter in the Universe, which has been determined precisely from 5 year WMAP data \cite{WMAP},
is given by $\Omega_{CDM} h^2 = 0.1099\pm 0.0062$. Assuming the coexistence of two dark matter candidates,
the relic abundance observed must
be composed of the contributions of both $S$ and $\psi$,
$\Omega_{_S} h^2 + \Omega_\psi h^2 = \Omega_{\rm CDM} h^2 $.
The relic density of each dark matter species is approximately given by
$\Omega_i h^2\approx (0.1 pb)/ <\sigma v>_i~~~~~~~(i=s,\psi),$
where $<\sigma v>_i$ is the thermally averaged product of its annihilation cross section with its velocity.
For our convenience, we define the parameter $\epsilon_i$ as a ratio of $\Omega_i h^2$ to
$\Omega_{\rm CDM} h^2$,
\be
\epsilon_i = \frac{\Omega_i h^2}{\Omega_{\rm CDM} h^2}~,
\ee
where $\epsilon_{_S}+\epsilon_{\psi}=1$.
In fact, the parameter $\epsilon_i$ represents the fraction of the mass density of each dark matter
species in our local dark-matter halo as well as in the Universe.
Since the values of $\epsilon_i$ are unknown, we consider a few cases by choosing their values in the analysis.
Each $\Omega_i h^2$ can be calculate with the help of the \emph{micrOMEGA}s 2.0.7 program \cite{micro} by taking input
parameters appropriately.

Except for the SM parameters, our model contains 18 new parameters: 6 scalar couplings $\lambda$, $\lambda_s$, $\lambda_{i(=1-4)}$,
4 masses of singlet particles $M_{\phi, \psi, S, N}$, 3 Yukawa couplings $Y_{D, S, \Phi}$, and 5 other parameters,
$\tan\theta, ~~v_{\phi, h},~~\epsilon_{S, \psi}$.
Among them, $M_N$, $Y_D$ and $Y_S$  are closely associated with low scale leptogenesis and light neutrino mass spectrum.
There also exist 7 conditions with which parameters should be satisfied, $e.g.$ $\epsilon_S + \epsilon_\psi =1$,
$cv_h - sv_\phi=v_{EW} = 246$ GeV, etc., and that
the parameters $\lambda$, $\lambda_i$ and $v_{\phi, h}$ are correlated with mass parameters for $h$ and $\phi$ given in Eq. (3).
 Accordingly, we have 8 free parameters: the parameters $m_{S,h,\phi}$, $\lambda_2$, $\tan\theta$
and $\epsilon_S$ (or $\epsilon_\psi$) are fixed by hand and
$\lambda~ ({\rm or}~ \lambda_3)$ is determined by the conditions.
(Another free parameter $\lambda_s$ is irrelevant in our analysis.)
In our numerical analysis, we take $\epsilon_{\psi}$
as an input parameter and then $\epsilon_{S}$ is
determined from the former relations and conditions.
Since both $\lambda$ and $\lambda_3$ are related to the $\epsilon_\psi$ parameter,
we can take either $\lambda_3$ or $\lambda$  as an input parameter and then the other one is
determined from the correlation among $\lambda$, $\lambda_3$ and $\epsilon_\psi$.
\\

\noindent{\bf II. Implication for the low mass DM signals:}
\\

In order to  interpret the low mass DM signals in terms of DM-nucleon scattering,
we choose $\psi$, the lighter DM particle of 2DM, to be relevant for the experiment. Note that
the heavier DM $S$ of order a few TeV is also demanded in order
to explain the high energy cosmic ray anomalies later.
To investigate the implication for the DM signals observed from the direct detections,
we first have to estimate the DM-nucleon elastic scattering cross section predicted in our scenario.
So far most experimental limits of the direct detections have been given in terms of the scattering
cross section per nucleon under the assumption that there exists only one DM candidate.
In the scenario of 2DM, the cross section for the WIMP-nucleon elastic scattering $\sigma_{el}$ is
composed of  $\sigma_{_S}$ and $\sigma_{\psi}$ \cite{ma};
\be \frac{\sigma_{el}}{m_0} = \frac{\epsilon_{_S}}{m_{_S}}\sigma_{_S} + \frac{\epsilon_\psi}{m_\psi}\sigma_\psi,
\label{sigma}
\ee
with $m_0$  being the WIMP mass, where we set $m_0=m_\psi$ as the relevant DM mass for direct searches.

\fig [tb]
\ct{\ep{figure=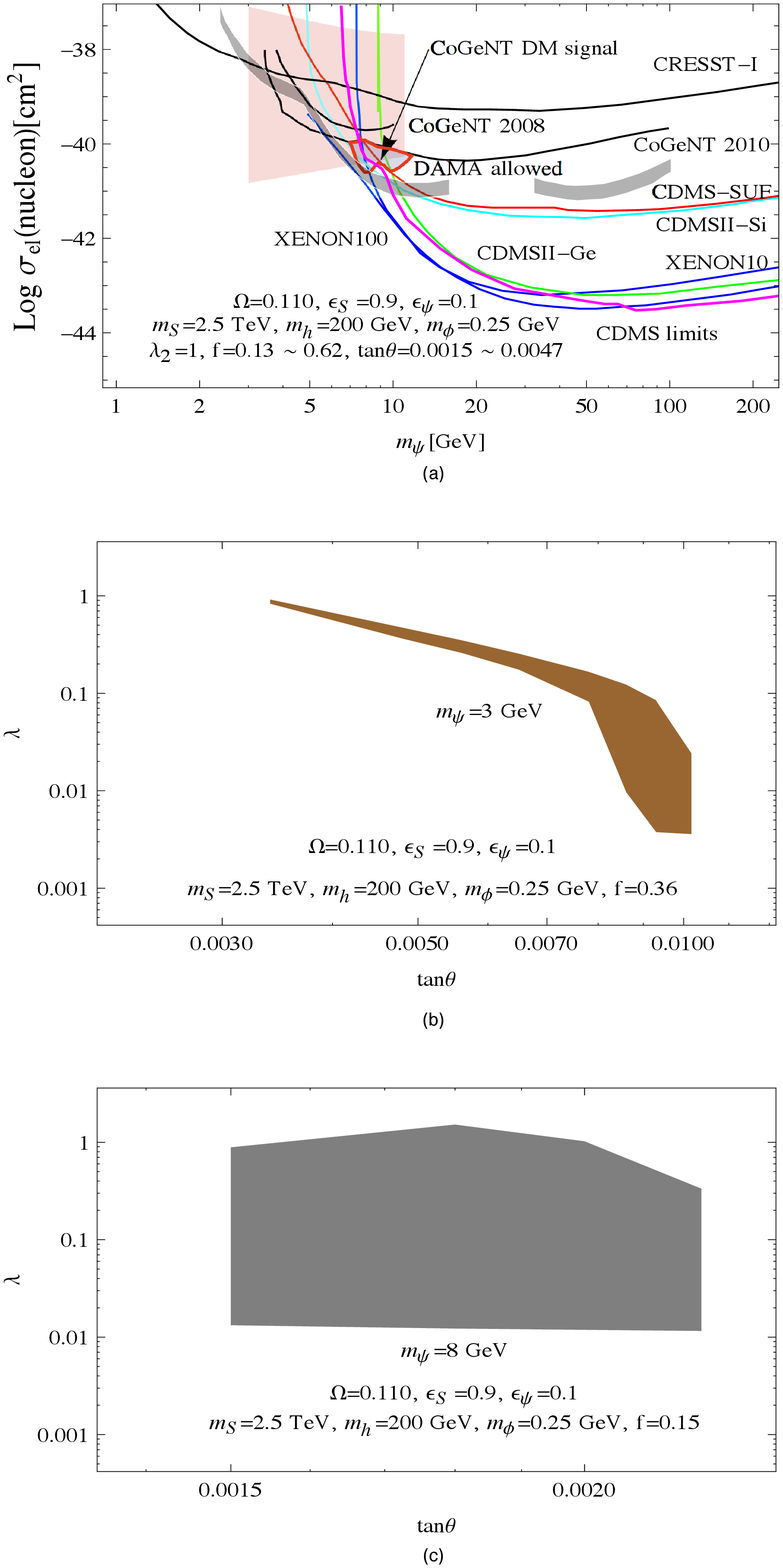, scale=0.45}}
\cp{(a)DM-nucleon
elastic scattering cross section $vs.$ DM mass. The DAMA results are
presented by the grey-colored regions, which includes both channeled
and quenched events as well as the allowed region consistent with
the DAMA modulation signal at $3\sigma$ \cite{petriello}. (In
fact, the recent estimates of the channeling effect shows that it is
smaller than expected \cite{bozorgnia}.) Also the pink-colored
rectangular region  corresponds to the prediction of our scenario
for given input values presented in the panel and
$3~\mbox{GeV}\lesssim m_{\psi}\lesssim 11~\mbox{GeV}$. The red
contoured region represents the DM signal from CoGeNT. (b) Allowed region of
the parameter space $(\tan\theta, \lambda)$ from the fit to the DAMA
results combined with the other null experiments for $m_\psi=$ 3
GeV, $f=$ 0.36 and the same input parameters as in (a). (c) Allowed
region from the fit to the results for DM signal from CoGeNT for
$m_\psi=$ 8 GeV, $f=$ 0.15 } \label{DAMAR} \ef

In our model, the non-relativistic $S$-nucleon elastic scattering cross section is given by
\be
\sigma_{_S} ({\rm nucleon}) \approx \frac{1}{4\pi}\Big[\frac{\sin 2\theta Y_\phi m_{_S} m^2_n f}{(m_n+m_{_S})v_h} \Big]^2
\Big [ \frac{1}{m^4_h}  + \frac{1}{m^4_\phi}  \Big ], \label{el1}
\ee
where $m_n$ is a nucleon mass and $f$ is defined by the relation
$f m_n \equiv <n|\sum_q m_q \bar{q} q |n>$ whose size is determined by \cite{scopel},
$ 0.13 \lesssim f \lesssim  0.62$. 
The first and second terms in the parenthesis correspond to the elastic scattering mediated by
the Higgs field $h$ and $SU(2)_L$ singlet scalar field $\phi$, respectively.
In the case of scalar $\psi$-nucleon elastic scattering,
the non-relativistic elastic scattering cross section for $\psi$ is given by
\be
\sigma_{\psi} ({\rm nucleon}) \approx \frac{1}{4\pi}\Big [\frac{m^2_n f}{(m_n + m_\psi)v_h} \Big ]^2
\Big [ \Big (\frac{\lambda^{\prime}c }{m^2_h} \Big )^2
+ \Big ( \frac{ \lambda^{\prime\prime}s}{m^2 _\phi} \Big )^2 \Big ],
\label{phielastic2}
\ee
where $\lambda^{\prime}=\lambda v_h c + 2\lambda_3  v_\phi s$
and $\lambda^{\prime \prime}=-\lambda v_h s + 2 \lambda_3 v_\phi c $.

In Fig. \ref{DAMAR}-(a), the pink-colored rectangular area presents the predicted region of the parameter space
($\sigma_{el} -m_{\psi}$) in our model
for several fixed input parameters given in the panel.
Here, we restricted the region of $m_{\psi}$ to be $3~\mbox{GeV} \lesssim m_{\psi} \lesssim 11~\mbox{GeV}$.
We see that  DAMA experimental result is consistent with other null experimental results
including CoGeNT 2010 (ignoring DM signal) and XENON100 data
only for the narrow range $m_\psi \sim 3 \mbox{ GeV}$.
We also see that our predicted region for DM mass range, $7 \mbox{ GeV} \lsim m_\psi \lsim 11 \mbox{ GeV}$,
is consistent with the DM signal observed from CoGeNT which corresponds to the red contour in Fig. \ref{DAMAR}-(a).
Fig. \ref{DAMAR}-(b) represents the allowed regions of the parameter space
$(\tan\theta, \lambda)$ from the fit to the DAMA results combined
with the other null results of direct searches particularly for $m_\psi=$ 3 GeV, $ f =$ 0.36 and the same input parameters as in Fig. \ref{DAMAR}-(a).
Fig. \ref{DAMAR}-(c) represents the allowed parameter region from the fit to the results of DM signals from CoGeNT for $m_\psi=8$ GeV and $f=0.15$.
When we calculate numerically scattering cross sections, we
vary $\lambda_3$ and $\tan \theta$ for a fixed value of $\epsilon_\psi$ as well as $\lambda_{s,2}$
and other mass parameters of the singlet particles including Higgs boson. And then the value of $\lambda$,
which lead to the right values of the scattering cross sections, can determined accordingly.
From our numerical calculation, we found that the lowest value of $\lambda$ is 0.01 which corresponds to $\lambda_3=0$.
The allowed values of $\lambda$ increases with $\lambda_3$, but there exists the upper bound on $\lambda$
for which the scattering cross section reaches the maximally allowed value for the DM signal from CoGeNT,
as can be seen in Fig. \ref{DAMAR}-(c).
We also notice that the excluded region $\tan\theta > 0.0022$ is not consistent
with the CoGeNT DM signal because the DM-nucleon cross section size can be larger
than the CoGeNT upper limit in the red contoured region. On the other hand, $\tan\theta < 0.0015$
is not consistent with electroweak symmetry breaking and relevant mass scale of the singlet scalar field $\phi$.
\\

\noindent{\bf III. Implication for PAMELA, ATIC and Fermi-LAT:}
\\

Now, let us show that the PAMELA, ATIC and Fermi-LAT data can be accounted for
by regarding singlet fermion $S$ as a relevant dark matter much heavier
than $\psi$, which annihilates into dominantly $\phi \phi$, and then the $\phi$'s subsequently decay into mostly $\mu^+ \mu^-$ when $m_{\phi}$
is taken to be 0.25 GeV.
In order to calculate the galactic cosmic ray (CR) propagation, we use GALPROP program
\cite{GALPROP} which simulates the propagation of both cosmic rays
and DM annihilation products in the galaxy. The propagation equation for all CR species is given in \cite{GALPROP}.
To solve the propagation equation under the assumption of free escape of particles at the halo boundaries,
we used the values of the parameters, which are based on the conventional model with constant Xco-factor
provided in the source code, galdef\_50p\_599278, placed in GALPROP web page \cite{GALPROPH}.
We normalized the primary electron flux to
 $3.2\times 10^{-10}~ cm^{-2} sr^{-1} s^{-1}\mkb{MeV}^{-1}$ at $34.5$ GeV
 so that it gives a good description of the data in our analysis.
 If the normalized electron flux is shifted, the background flux of positron fraction
 is also changed so as to make it difficult to fit all the data points of PAMELA.
In addition, we use an NFW density profile \cite{NFW}, so that the core radius and the local DM density are taken to be $20.0$ kpc
and $0.3~\mkb{GeV} cm^{-3}$, respectively.

\fig [h] \ct{\ep{figure=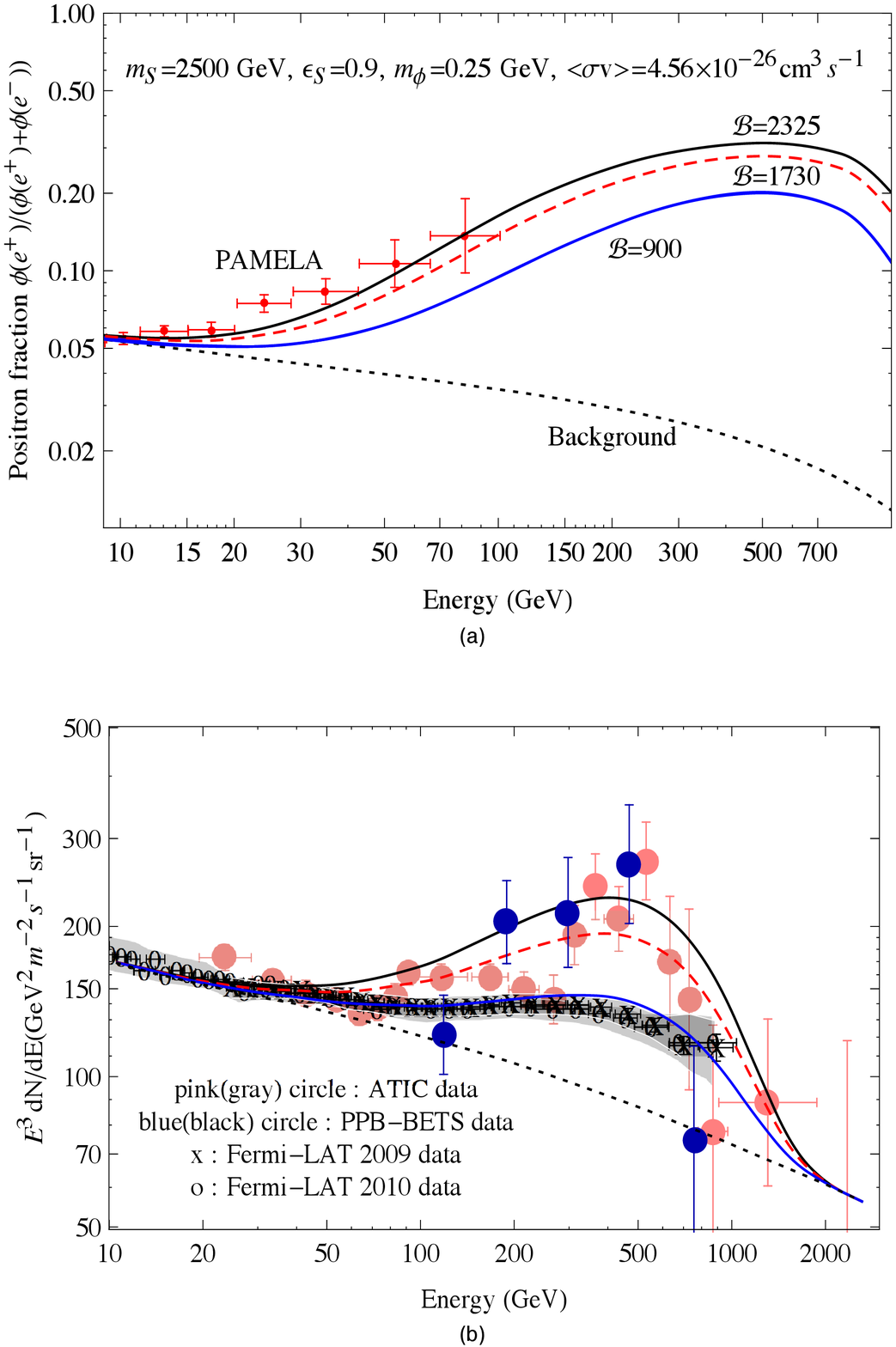, scale=0.45}}\cp{(a) The ratio of positron to electron plus positron fluxes
and (b) total electron plus positron fluxes,  arising from the annihilations $SS\rightarrow \phi\phi$
and then $\phi\rightarrow \mu^+ \mu^-$. ${\cal B}$ stands for the boost factor relative to $<\sigma v> = 4.56 \times 10^{-26}~cm^3 s^{-1}$
which is satisfied thermal relic abundance for $\epsilon_{_S} = 0.9$  \label{PA}}
\ef

In Fig. \ref{PA}, we present the predictions of our scenario for
(a) the ratio of positron to electron plus positron fluxes
and (b) the total electron plus positron fluxes, which are originated from $SS\rightarrow \phi\phi$,
and subsequent decays of the $\phi$'s into $\mu^+ \mu^-$ for the same input values of the parameters.
As for input values, we take $\epsilon_{_S} = 0.9$, $m_{_S}=2.5$ TeV, $m_\phi =0.25$ GeV, and $<\sigma v>=4.56\times 10^{-26}cm^3 s^{-1}$
which satisfies the thermal relic density of $S$ for the given $\epsilon_{_S}$.
Then, the contribution of $S$ to the local DM density ($\rho^0_{_S}$ ) is $0.27~ \mkb{GeV } cm^{-3}$
while that of $\psi$ to the local DM density ($\rho^0 _\psi$) is $0.03~ \mkb{GeV } cm^{-3}$.
In these estimates, we invoke the boost factor (${\cal B}$) reflecting
Sommerfeld enhancement through which  the halo annihilation rate is enhanced,
and it is given by
\be
{\cal B} \sim \frac{\al m_{_S}}{m_\phi},
\ee
where $\alpha$ lies between $10^{-3}$ and $ 10^{-1}$ \cite{Arkani}.
The each curve in Fig. \ref{PA} corresponds to different boost factor, ${\cal B}$=2325 (1730, 900)
for black solid (red dashed, blue solid) curve.
The red dots with error bar correspond to the measurements from the PAMELA (Fig. \ref{PA}-(a)).
The pink (grey), blue (black) circles, and the points denoted by ``x" and ``o"  in Fig. \ref{PA}-(b) correspond to the measurements
from ATIC, PPB-BETS, Fermi-LAT 2009 and Fermi-LAT 2010, respectively.
As one can see from Fig. \ref{PA}-(a),
the black solid and red dashed curves corresponding to ${\cal B}$=2325 and ${\cal B}$=1730 give acceptable fits
to the PAMELA data for the positron fraction.
For the same values of ${\cal B}$, as in Fig. \ref{PA}-(a), the predictions of $E^3dN/dE$ as a function
of $E$ appear to give acceptable fits to the ATIC and the PPB-BETS data
as well as the Fermi-LAT data for $E\lesssim 70$ GeV, whereas the predictions for $70\lesssim E \lesssim 750$ GeV
are much deviated from the Fermi-LAT data .
On the other hand, we see that the prediction of  $E^3dN/dE$ for ${\cal B}$=900 (blue curve in Fig. \ref{PA}-(b))
gives acceptable fit to the Fermi-LAT data, but that of positron fraction is quite small
 to fit well the PAMELA data as shown in Fig. \ref{PA}-(a).
Therefore, it looks rather difficult to perfectly accommodate the PAMELA, ATIC and Fermi-LAT data
simultaneously, which may imply that there exist other astronomical sources \cite{pulsar, supernovae, Malyshev, Choi}.
In passing, please note that the introduction of one more generation of singlet vector-like neutrino, $S$,
slightly weakens the tension between the cosmic ray data and allow for lower boost factor (even below 700).
However the model loses some of its predictive power due to several new additional parameters.
\\

In conclusion, we have shown that the extended seesaw model proposed to achieve low scale leptogenesis can
resolve the anomalies in the indirect detections of annihilation products observed from PAMELA, ATIC
and/or Fermi-LAT and {\it simultaneously} accommodate  some of recent signals of low mass DM measured at DAMA and CoGeNT.
In this model, an extra vector-like singlet neutrino $S$ and a singlet light scalar field $\psi$,
which are coexisting two-particle dark matter candidates,
are responsible for the origin of the excess positron and electron fluxes and
the low mass DM signals observed from DAMA and CoGeNT.
Furthermore, it has been shown that the DM signal observed from DAMA  and the other null results
including CoGeNT 2010 and XENON100 data
from direct searches for DM can be reconciled in the case of $m_{\psi} \sim 3~\mbox{GeV}$.
We have also shown that the DM signals observed from CoGeNT can be accommodated in our model
if $7~\mbox{GeV} \lesssim m_{\psi} \lesssim 11~\mbox{GeV}$.
On the other hand, in addition to $SU(2)_L$ doublet Higgs field $H$, the (light) singlet Higgs field $\Phi$, which
is demanded to successfully construct the coexisting two-particle dark matter scenario
and whose mass is taken to be just below 1 GeV,
may play an essential role in resolving the PAMELA, ATIC and/or Fermi-LAT anomalies.

\newpage
\noindent{\bf Acknowledgements:}
\\
C.S.K. and H.S.C. are supported in part by Basic Science Research Program through
the NRF of Korea
funded by MoEST (2009-0088395) and in part by KOSEF through the Joint Research Program (F01-2009-000-10031-0).
S.K.K. is supported in part by the Korea Research Foundation (KRF)
grant funded by the Korea government(MoEST) (2009-0090848).
\\

\end{document}